%% file: Allerton2023_main.tex
\documentclass[conference,twocolumn]{IEEEtran}

\usepackage{xcolor}
\usepackage{soul}
\usepackage{multirow} 
\usepackage{cite}
\usepackage{relsize}
\usepackage[labelsep=period]{caption}
\usepackage[left=0.6in,right=0.6in,top=0.7in]{geometry}

\usepackage{amsmath}
\usepackage{varwidth}
\usepackage{amssymb}
\usepackage{algorithm}
\usepackage{algpseudocode}
\usepackage{mathtools}
\usepackage[cmintegrals]{newtxmath}
\usepackage{array}
\usepackage[caption=false,font=normalsize,labelfont=sf,textfont=sf]{subfig}

\usepackage{fixltx2e}
\usepackage{stfloats}
\usepackage{enumitem}

\begin{document}

\title{An Efficient Distributed Multi-Agent Reinforcement Learning for EV Charging Network Control \\
}


\author{\IEEEauthorblockN{Amin Shojaeighadikolaei, Morteza Hashemi}

\IEEEauthorblockA{Department of Electrical Engineering and Computer Science, University of Kansas, Lawrence, KS, USA}

}

\maketitle

\begin{abstract}
The increasing trend in adopting electric vehicles (EVs) will significantly impact the residential electricity demand, which results in an increased risk of transformer overload in the distribution grid. To mitigate such risks, there are urgent needs to develop effective EV charging controllers. Currently, the majority of the EV charge controllers are based on a centralized approach for managing individual EVs or a group of EVs. In this paper, we introduce a decentralized Multi-agent Reinforcement Learning (MARL) charging framework that prioritizes the preservation of privacy for EV owners. We employ the Centralized Training Decentralized Execution-Deep Deterministic Policy Gradient (CTDE-DDPG) scheme, which provides valuable information to users during training while maintaining privacy during execution. Our results demonstrate that the CTDE framework improves the performance of the charging network by reducing the network costs. Moreover, we show that the Peak-to-Average Ratio (PAR) of the total demand is reduced, which, in turn, reduces the risk of transformer overload during the peak hours. 
\end{abstract}


\input{Conference_paper/Chapters/Introduction_Version3}

\section{Literature Review}\label{LitReview}
A growing body of literature studies have examined different model-free RL deployment to address EV charging control and EV scheduling problems. For example, the authors in~\cite{wen2015optimal} have formulated the EV charging problem as an MDP and utilized the Q-learning with a lookup table to learn the best charging strategy. In 2017, Chis \textit{et al.}~\cite{chics2016reinforcement} proposed a demand response method that aims at reducing the long-term cost of charging of an individual plug-in electric vehicle by using Bayesian neural network for predicting the electricity price. Similarly, in~\cite{wan2018model}, the authors utilized two networks, including a representation network to extract discriminative features from the electricity prices and a Q-network to approximate the optimal charging/discharging power for an individual EV. Inspired by these works, Zhang \textit{et al.}~\cite{zhang2020cddpg} leveraged an advanced RL method known as Deep Deterministic Policy Gradient (DDPG) for charging control of an individual EV. They utilized a time series long short-term memory (LSTM) network to extract the previous energy price information. As an extension of  the single-EV charging control, the authors in~\cite{jin2020optimal} studied the scheduling of large-scale electric vehicles charging in a power distribution network under random renewable generation and electricity prices by using soft-actor-critic (SAC) algorithm. 

The aforementioned single-agent approaches assume full observability of the environment and train a single agent to control either a single EV or a group of EVs in a centralized manner. However, this assumption raises privacy concerns when controlling a network of EVs, as the single agent becomes a potential point of attack due to its possession of all the corresponding information. Additionally, as the number of EVs within the network increases, it becomes challenging for a single agent to handle the network requirements effectively. To tackle this challenge, MARL emerges as a solution for such scenarios. MARL has been applied in many applications, e.g., Volt-VAR control in distribution networks~\cite{gao2021consensus}, microgrid energy management~\cite{fang2021multi}, and traffic flow control~\cite{zeynivand2022traffic}. For EV networks, the authors in~\cite{da2019coordination} proposed a multi-agent collaborative architecture to minimize the energy costs to avoid transformer overloads during charging phase. In this work, they consider time-of-used price tariffs. Chu \textit{et al.} in~\cite{chu2022multiagent} and Zhang \textit{et al.} in~\cite{zhang2022federated} utilized federated reinforcement learning to solve the problem in a distributed manner. In~\cite{zhang2023cooperative}, the authors investigated EV charging/discharging collaborative control based on DDQN for an EV station. In one of the most recent works, Yan \textit{et al.} in~\cite{yan2022cooperative}, proposed a cooperative MARL framework for charging control of residential EVs by constructing a DNN network to approximate the behaviors of other agents and independent learners using soft-actor-critic method. In contrast to~\cite{yan2022cooperative} we relax the assumption that each EV is capable of observing the total network demand at any given time. 

In this paper, we outline our method for modeling and controlling of EVs charging in a residential network. We utilize a POMDP framework and incorporate the advanced MADDPG technique. Unlike previous methods, we adopt an CTDE architecture, utilizing additional information for training purposes. The only shared observation among agents is the electricity price. During execution, local actors enable effective and decentralized control over the charging problem by utilizing the local information.

\input{Conference_paper/Chapters/System_model2}

\section{Distributed Algorithm}\label{Algorithm}
\subsection{Pricing tariffs}
In electricity marketplace, electricity price is the only signal that is observable by all the users through the network. In this paper, we define a function $\mathbf{F}_h(L_h)$ indicating the electricity price, which is a function of the network total consumption at time step $h$. In particular, we make the following assumption throughout this paper:

\textit{Assumption1:} The price function is increasing. That means, for each $h \in \mathcal{H}$, the following inequality holds:
\begin{equation}
    \mathbf{F}_h(\Tilde{L}_h) < \mathbf{F}_h(L_h)   \ \ \ \ \forall \ \Tilde{L}_h < L_h
\end{equation}

\textit{Assumption2:} The price function is strictly convex. That is, for each $h \in \mathcal{H}$, any real number $L_h,\Tilde{L}_h \geq 0$,and any real number $0 < \theta < 1$, we have
\begin{equation}
    \mathbf{F}_h(\theta L_h + (1-\theta)\Tilde{L}_h) < \theta\mathbf{F}_h(L_h) + (1-\theta)\mathbf{F}_h(\Tilde{L}_h)
\end{equation}

An interesting example for such electricity price function that satisfy aforementioned assumptions is the quadratic function. In this paper, we consider this function as follows:
\begin{equation}
    \mathbf{F}_h(L_h) = a_h (L_h)^2 + b_h L_h + c_h,
\end{equation}
where $a_h>0$ and $b_h, c_h\geq 0$ at each time step $h$. The electricity price is a function of the network's demand, and this paper assumes that users do not have knowledge of the underlying function. Instead, they only have access to periodic samples of the electricity price, without any prior information about how the price function operates. Therefore, in order to better interact with the network, users need to learn the price function. To this end, RL is useful that can help the users to learn the price function based on the samples during the time. 

\subsection{Principle of the Algorithm}
Consider  any user $i \in \mathscr{V}$. We consider each user as an RL agent, with each agent having partial observability of the network environment. This type of problem is commonly referred to as a MARL problem. The problem of MARL is formulated as a Decentralized POMDP (Dec-POMDP) \cite{oliehoek2016concise} like a tuple, such that $ \langle \mathcal{I}, \mathcal{S}, \mathcal{A}, \mathcal{O}, \mathcal{T}, \mathcal{R}, \gamma \rangle$. $\mathcal{I}$ is the agent set. In this paper, $\mathcal{I} \equiv \mathscr{V}$ ; $\mathcal{S}$ is a state set; $\mathcal{A}$ is  a joint action set; $\mathcal{O}$ is a joint observation set, where $\mathcal{O}_i$ is each agent's observation set; $\mathcal{T}: \mathcal{S} \times \mathcal{A} \times \mathcal{S} \rightarrow [0,1]$ is a transition probability function; $\mathcal{R}:\mathcal{S} \times \mathcal{A} \rightarrow \mathbb{R}$ is the reward function set, and $\gamma \in (0,1)$ is a discount factor. Each agent aims to maximize its own total expected reward, as $\mathbb{E}[R_i] = \max_\pi \mathbb{E}_\pi [\sum\limits_{t=0}^{T} {\gamma^t r_i^t}]$ where $r_i^t \in \mathcal{R}$ is the collected reward by the $i^{th}$ agent at time $t$ and $T$ is the time horizon.

To implement the distributed algorithm, it is necessary to refer to the system model presented in Figure~\ref{System Model}. Each user in this system is an RL agent and is equipped with an EV charging controller and a smart meter. The main goal of each agent is to maximize its individual profit. The environment in this problem is highly non-stationary from the perspective of individual agent~\cite{lowe2017multi}, since it is not only dependent on their actions but also on the joint action of all other agents. To address this issue, the simplest approach is to train each agent independently, but this approach ignores the non-stationarity of the environment. A more effective approach is the CTDE paradigm, which has become widely used recently to overcome the non-stationarity problem.
In this paper, we explore a cooperative multi-agent reinforcement learning (CMARL) problem for an EV charging network, where all agents operate under partial information. In our framework, the agents are DDPG agents and by leveraging CTDE, we allow the policies to use additional information to facilitate training, but this information is not used during testing. 

DDPG is a policy gradient algorithm, which consists of two networks: Actor and Critic, where the parameterized actor function $\mu(s\mid \theta^\mu)$, with parameter $\theta^\mu$, holds the policy and deterministically maps the states to a specific action. The term deterministic refers to the fact that the actor network output the exact output instead of the probability distribution over actions, $\mu(s) = \arg\max_a{Q(s,a)}$. In addition, the critic network describes as action-value function $Q(s,a|\theta^{Q})$, with parameter $\theta^{Q}$. The policy gradient algorithms are based on the idea that directly adjust the policy's parameter $\theta$ in the direction of $\nabla_{\theta}J(\theta)$ in order to maximize $J(\theta) = \mathbb{E}[R]$.

Consider the EV network with $\mathscr{N}$ agents with policy parameterized by $\boldsymbol{\theta} = \{\theta_1,..., \theta_N \}$ and let $\boldsymbol{\mu}=\{ \mu_1, ..., \mu_N \}$ be the set of Critics and deterministic policies, respectively. In order to prevent non-stationarity, MADDPG uses the actions and observations of all agents in the action-value functions and as the policy of an agent is only conditioned upon its own private observations, the agents can act in a decentralized manner during execution. Thus, the gradient of the expected return for the agent $i$ is defined as follows:
\begin{align}
    \nabla_{\theta_i} J(\mu_i) = \mathbb{E}_{o,a \sim \mathcal{D}} [\nabla_{\theta_i}\mu_i(a_i|o_i) \nabla_{a_i}Q_i^\mu (o_1,a_1,... \nonumber \\ ,o_N,a_N) |_{a_i=\mu_i(o_i)}].
\end{align}

Since each $Q_i^\mu$ is learned separately, agents can have different reward structure. Here, the experience replay buffer $\mathcal{D}$ contains the tuple $\langle o, o^\prime, a_1,...,a_N,r_1,...,r_N \rangle$. The centralized action-value function $Q_i^\mu$ is updated as:
\begin{equation}
    \mathcal{L}(\theta_i) = \mathbb{E}_{o,a,r,o^\prime}[(Q_i^\mu(o_1,a_1,...,o_N,a_N)-y)^2],
\end{equation}

where $y=r_i + \gamma Q_i^{\mu^\prime}(o_1^\prime, a_1^\prime,...,o_N^\prime, a_N^\prime)|_{a_j^\prime = \mu_j^\prime(o_j^\prime)}$, and $\boldsymbol{\mu}^\prime = \{\mu_{\theta_1^\prime},..., \mu_{\theta_N^\prime} \} $ is the set of target policies.
\vspace{0.1in}

\textbf{Action Set}: Each agent $i \in \mathscr{V}$ is equipped with a continuous action set $\mathcal{A}_i = \{a_i: 0\leq a_i \leq a^{max}, a^{max} > 0 \}$. The continuous action represents the charging volume. The EV battery level is calculated as $B_i(h+1) = B_i(h) + (\eta \times a_i^h \times \Delta t )/C$, where $B_i(h)$ is the battery level at time $h$, $\eta$ is the battery efficiency, $a_i^h$ is the charging power rate in $kW$, and $C$ is the battery capacity in $kWh$. $\Delta t$ is the charging period in which the charging rate is constant. 
\vspace{0.1in}

\textbf{Observation Set}: At any given time $h$, each agent $i \in \mathscr{V}$, has a partial observability of the environment. During the training phase, the observation set for each agent $i$ is defined as
$o_i = \{ \Delta B_i, \Delta h_i, \mathbf{F}_h, Pl_i, h_i^{dep} \}$, where $\Delta B_i = B_i^{exp} - B_i(h)$ is the difference between the amount of desired battery level and the current battery level at any given time. $\Delta h_i = h - h_i^{arr}$ represents the difference between arrival time $h_i^{arr}$ and current time step. $\mathbf{F}_h$ is the electricity price at time step $h$. $Pl_i$ is a binary flag. $Pl_i=0$ represents the corresponding EV of agent $i$ is not connected to the charging network and $Pl=1$ represents the corresponding EV is plug-in, and $h_i^{dep}$ denotes the departure time of the corresponding EV of agent $i$. 

\textbf{Reward Function}: 
In MADDPG, each agent has its own reward function, which is based on the local observation of the agent. $r_i^h$ represents the immediate reward for the agent $i$, at time $h$, which is obtained after the state changes from $s_i^h$ to $s_i^{h+1}$ by executing action $a_i^h$. According to the objective of the user satisfaction and network requirements, the reward function is defined as follows:
\begin{equation}
    r_i^h = - \alpha_1 \times \mathbf{F}_h \times a_i^h - \alpha_2 \times (B_i(h) - B_i^{exp})^2 + \mathcal{E},
\end{equation}
where, $\alpha_1$ and $\alpha_2$ are constant coefficient and $\mathcal{E}$ is a penalty term provides a large negative reward based on the distance from the expected battery level, if the agent loose to fully charge the EV at the departure time.

\section{Numerical Results}\label{results}
In this section, we evaluate the performance of our proposed framework of cooperative DDPG for EV charging network. To illustrate the performance of our framework, we consider a non-cooperative MADDPG framework, where each agent observes the local information and there is no direct information exchange even during the training phase. We call this scheme as I-DDPG. On the other hand, hereinafter, we call our proposed framework as CTDE-DDPG scheme. In this section, we demonstrate that CTDE-DDPG improves the performance of I-DDPG algorithm. We first present the experimental setting and then depict the effect of networked MARL on EV charging performance. We also demonstrate the convergence process of CTDE-DDPG and compare it with I-DDPG.

\subsection{Simulation Settings}
Our simulation results divided into two parts: \textbf{(1)} to demonstrate the general performance of the framework considering different EV owner behavior, we simulate the EV network charging control based on Table~\ref{RandomVariables}, where arrival time, departure time, the expected battery level at departure time, and the amount of battery energy at arrival time, all obey limited normal distribution. (2) To better demonstrate the cooperation and charging scheduling between agents, in part two we fix the arrival time, departure time, EV energy at arrival time, and the battery capacity and compare the CTDE-DDPG and I-DDPG scenarios.

\subsection{Part 1: General Performance}
In this simulation, inspired by~\cite{zhang2020cddpg}, the charging behavior is modeled randomly using Table~\ref{RandomVariables}. More specifically, the arrival time is sampled as an integer from a normal distribution with a mean of 9 and a standard deviation of 1, subject to the constraint of $7 \leq h^{arr} \leq 12$. similarly, the departure time is sampled from a normal distribution with a mean of 18 and a standard deviation of 1 with a limitation of $\mathcal{N}(18,1^2)$. In addition, we sample desired battery level at departure time and initial battery level at arrival time from $\mathcal{N}(55,1^2)$ and $\mathcal{N}(5.5,1^2)$ within the intervals $[45,65]$ and $[4.5,6]$, respectively.
\begin{figure*}[!t]
   \centering
    \includegraphics[scale=0.09, trim = 0.2cm 0.2cm 0.2cm 0.2cm, clip]{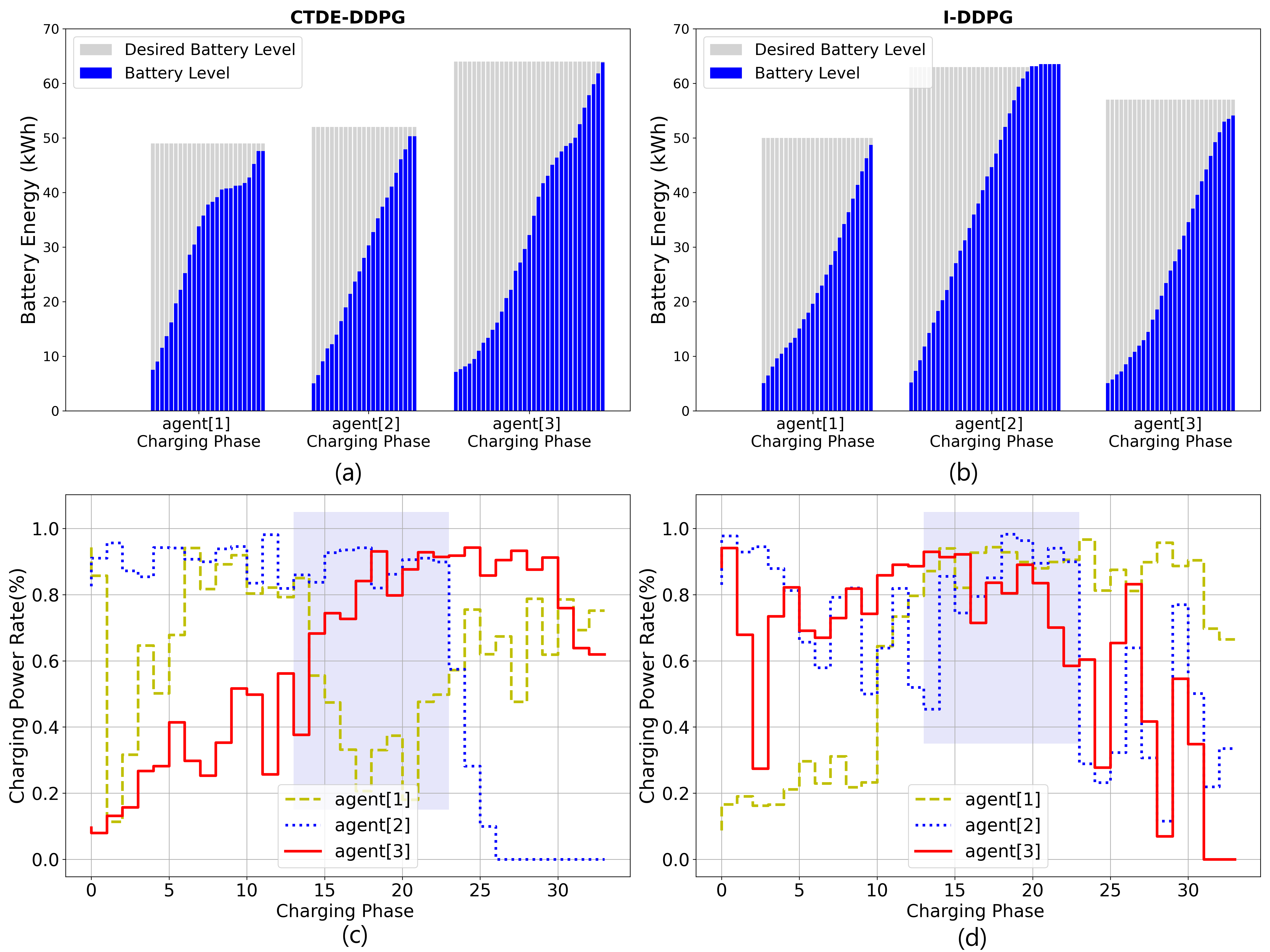}
    \caption{(a) - (b) Performance comparison for CTDE-DDPG and I-DDPG schemes in terms of battery level during the charging phase. (c)-(d) A charging behavior comparison between CTDE and independent DDPG over a charging phase.}
    \label{3agents_behavior}
    \vspace{-.1in}
\end{figure*}
\begin{figure}[!t]
   \centering
    \includegraphics[scale=0.47, trim = 0.2cm 0.2cm 0.2cm 0.2cm, clip]{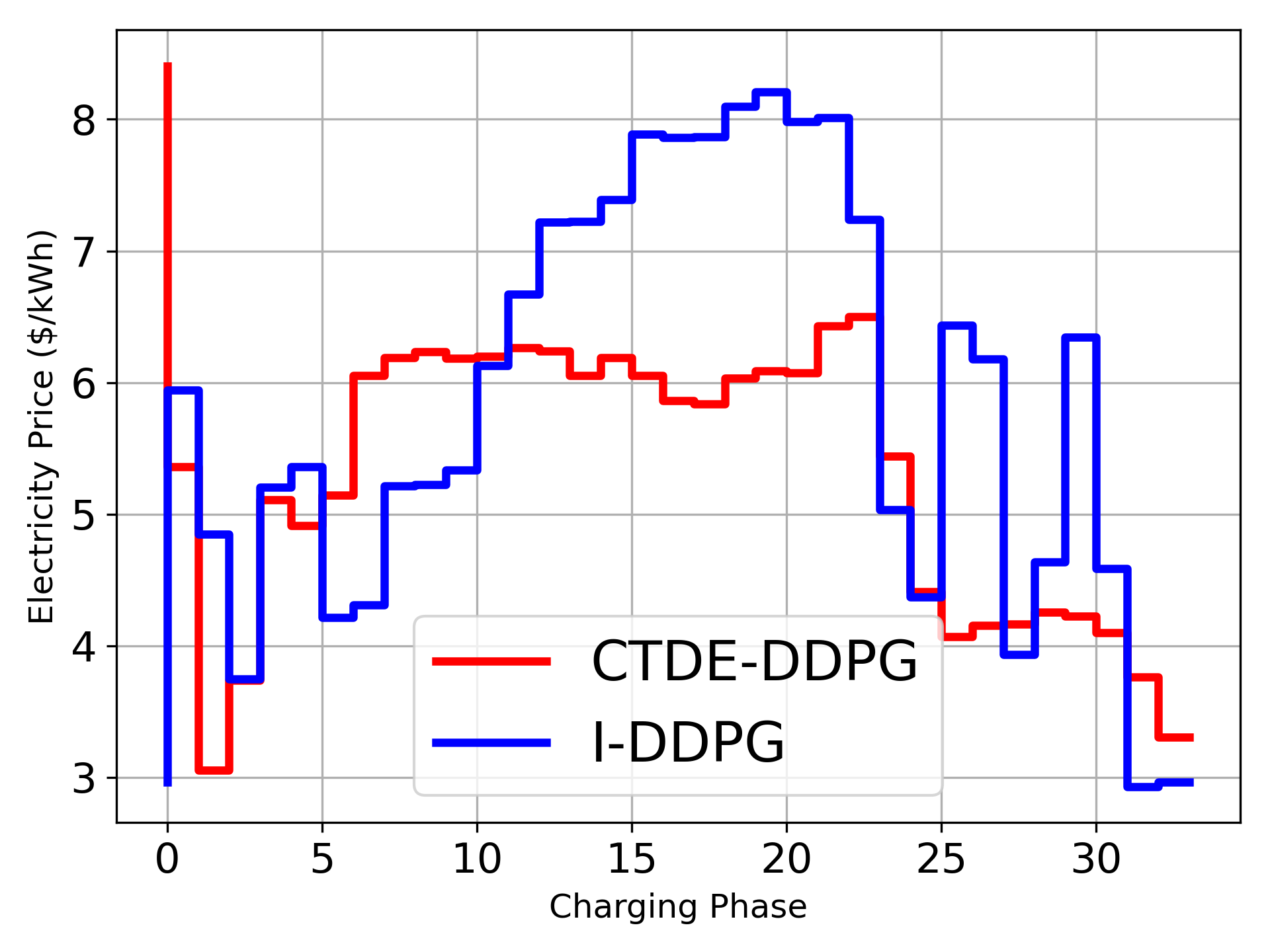}
    \caption{Average electricity price comparison of CTDE-DDPG and I-DDPG schemes.}
    \label{Avgprice}
    \vspace{-.1in}
\end{figure}
\begin{table}[!t]
\caption{RANDOM VARIABLES FOR MODEL PARAMETERS IN SIMULATION RESULTS.}
\resizebox{\columnwidth}{!}{
\centering
\footnotesize
  {
    \begin{tabular}{l|l|l}
      \hline
      \textbf{Model Parameters} & \textbf{Distribution} & \textbf{Limits}\\
      \hline
      Arrival time & $\mathcal{N}(9,1^2)$ & $7 \leq h^{arr} \leq 12$ \\
      \hline
      Departure time & $\mathcal{N}(18,1^2)$& $16 \leq h^{dep} \leq 20$ \\
      \hline
      Expected battery level at departure time & $\mathcal{N}(55,1^2)$ & $45 \leq B^{exp} \leq 65$ \\
      \hline
      Battery level at arrival time & $\mathcal{N}(5.5,1^2)$ & $4.5 \leq B^{arr} \leq 6$ \\
      \hline
    \end{tabular}
    }}
 \label{RandomVariables}
  \vspace{-0.15in}
\end{table}
In the following, we investigate the general performance of the proposed CTDE-DDPG framework and compare it with I-DDPG. It is worth to mention that the primary goal of both algorithms is to fully charge the EVs at the departure time to meet the EV's owner's demand. Fig.~\ref{3agents_behavior} (a) and (b) compare the amount of battery energy of CTDE and independent DDPG over a specific day for three agents. The blue bar indicates the battery level, and the gray bars are the desired battery energy amount during the charging phase. As illustrated in Fig.~\ref{3agents_behavior} (a) and (b), the DDPG-based charging control could satisfy user's demands for battery energy in both scenarios. In both scenarios, the charging control successfully charges the EVs during the designated phase and fulfills the customers' expected battery level requirements. From the Fig~\ref{3agents_behavior} (a) and (b), we observe that the charging phase exhibits similar behavior among three agents. However, in the CTDE scenario, the charging behavior undergoes alterations due to different rates over time, implying that the agents do not exhibit identical charging behavior in this scenario.
\begin{figure*}[!t]
   \centering
    \includegraphics[scale=0.18, trim = 0.2cm 0.2cm 0.2cm 0.2cm, clip]{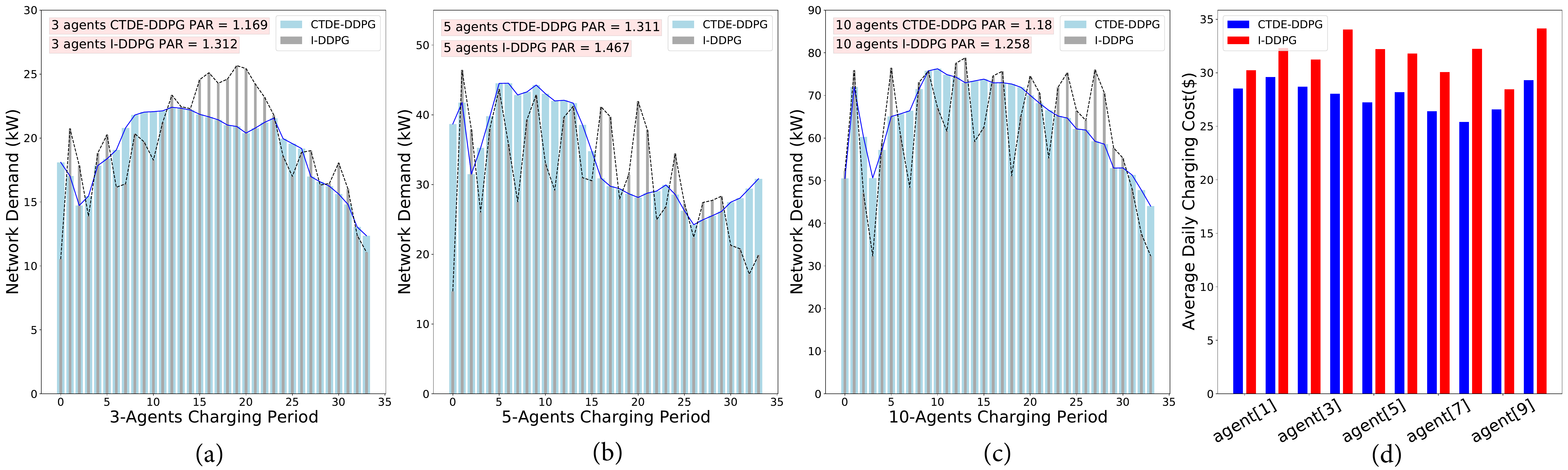}
    \caption{(a)-(c) Impact of number of the agents to the network performance in terms of PAR. (d) Average cost comparison during the charging phase for 10-agents scenario.}
    \label{PARandCOST}
    \vspace{-.1in}
\end{figure*}
To investigate this alteration in details, Fig.~\ref{3agents_behavior} (c) and (d) depict the comparison of average charging rate over the charging phase for 3-agent scenarios. In the I-DDPG scenario, the charging behavior reveals a consistent pattern among all three agents. Especially, during the highlighted areas, we observe simultaneous high-rate charging by all three agents in I-DDPG scenario. All three agents charge their batteries at full rate within the highlighted time and this resulting in an increased network demand during those instances. Conversely, in the CTDE scenario, the three agents strive to collaborate and charge their batteries at varying intervals throughout the charging phase. This results in consistency in electricity price and network total demand waveforms. As shown in Fig.~\ref{Avgprice}, in the I-DDPG scenario, when all three agents choose to charge their batteries simultaneously at a high power rate, the electricity price increases. This results in a financial disadvantage for all users during their charging phase. Alternatively, in the CTDE scenario, the electricity price remains consistent during the charging phase, highlighting the economic benefits of agents' cooperation. 


\subsection{Part 2: Cooperation vs. Scalability}
In this section, we evaluate the impact of increasing the number of agents in our system model on the performance of the proposed CTDE-DDPG framework. To better compare our framework with the I-DDPG, we compare the Peak-to-Average (PAR) performance of the two cases. PAR has been used extensively in the literature as a parameter to measure the effectiveness of the demand-side management algorithms, and is defined as follows:
\begin{equation}\label{eq:PAR}
\text{PAR} = \frac{T \max_{t \in \mathcal{T}}{L^D_t}}{\sum\limits_{t \in \mathcal{T}}{L^D_t}}.
\end{equation}

Fig.~\ref{PARandCOST} (a)-(c) provides three subplots, each presenting a comparison of EV network total demand between the I-DDPG and CTDE-DDPG scenarios. These subplots depict the numerical analysis for 3-agents, 5-agents, and 10-agents cases. In 3-agents case, the PAR for CTDE and Independent DDPG cases are 1.177 and 1.331 respectively. As expected, these results align with the findings from the previous section, where all three agents in the I-DDPG scenario exhibited similar behavior and occasionally charged their batteries simultaneously. This simultaneous charging lead to an increase in the total network demand and subsequently elevate the PAR. In emergency situations when the  network demand is high, the proliferation of charging EVs can potentially jeopardize grid stability. 

To investigate the impact of increasing the number of agents, Fig.~\ref{PARandCOST} (b) and (c) also presents the results for the 5-agent and 10-agent cases. In the 5-agent scenario, the PAR values for CTDE and Independent DDPG are 1.303 and 1.4858, respectively. For the 10-agent scenario, the corresponding PAR values are 1.162 for CTDE and 41.24 for Independent DDPG. As depicted, the network demand becomes smoother as the number of agents increases. However, it is worth noting that the difference between CTDE and Independent DDPG diminishes in terms of PAR.

Despite the similarity in network performance, as measured by PAR, between CTDE-DDPG and I-DDPG with an increasing number of agents, the individual performance of the agents varies. Fig.~\ref{PARandCOST} (d) provides a visual representation of this distinction by showcasing the average cost of agents in the 10-agent scenario. The results demonstrate that in the CTDE scenario, the local cost for all 10 agents is lower compared to the I-DDPG scenario.

\input{Conference_paper/Chapters/Conclusion}




\bibliographystyle{plain}
\bibliography{refs}

\end{document}

%% file: Conference_paper/Chapters/Introduction_Version3.tex
\section{Introduction}

Demand-side management (DSM) is a strategy employed by grid operators to effectively manage the total demand on an electricity network. The main objective of DSM is to decrease the overall electricity demand by encouraging consumers to shift their usage to non-peak hours or reduce their overall consumption. In traditional DSM programs, end-users interact directly with the utility company. However, in the modern smart grid, the interaction shifts towards grid operators engaging with a network of end-users, enabling end-users to communicate with each other.

The increasing adoption of Electric Vehicles (EVs) in end-users network, presents both opportunities and challenges for DSM initiatives. While EVs contribute to a sustainable energy system by reducing transportation carbon emissions, they pose significant challenges, such as increasing the energy demand during peak periods. To address this challenge, effective management and coordination of EV charging and discharging behavior are essential. The objective of EV charging control is to ensure a smooth and reliable charging experience while maximizing the benefits for both individual EV owners and the overall grid network. In terms of individual benefits, EV charging control aims to minimize charging costs for owners and meet their battery energy requirements, which has been extensively researched~\cite{al2019review, abdullah2021reinforcement}. Additionally, for the grid network, charging control seeks to balance charging demand and prevent transformer overload during peak demand hours, which still remains an area under investigation~\cite{nimalsiri2019survey}. 

However, designing an optimal charging strategy for a network of EVs is a complex task due to several challenges. One of the primary challenges is the uncertainty in dynamic electricity prices, which is currently one of the most intriguing areas of focus in the field of smart grid technologies. Dynamic pricing can result in price fluctuations, making it difficult for EV owners to accurately predict the cost of charging their vehicles. Another challenge arises from the EV owners' behavior, which can vary widely. The arrival and departure times of EVs to the charging network do not follow a specific distribution, and EV owners may have different preferences regarding their desired charging level and preferred charging time. However, one of the most crucial challenges in an EV network is congestion management and transformer overload, which is the main focus of this paper. Simultaneous charging EVs have the potential to overwhelm the transformers connected to the network, and an optimal charging strategy should effectively manage this congestion. In this case, coordination and communication between the EVs can control and adjust their charging behavior based on real-time grid conditions. 

Considering the aforementioned challenges, for avoiding transformer overload, researchers have developed many approaches to coordinate the charging process of EVs network. These approaches can basically be categorized into two groups based on their model assumption: model-based and model-free methods. Model-based approaches such as binary optimization~\cite{sun2016optimal}, mixed-integer linear programming~\cite{paterakis2016coordinated}, robust optimization~\cite{ortega2014optimal}, stochastic optimization~\cite{wu2017two}, model predictive control~\cite{zheng2018online}, and dynamic programming~\cite{xu2016dynamic} have been used for EV charging control and optimal scheduling considering electricity price uncertainty and users behavior.  In these approaches, the scheduling of EV charging is typically framed as an optimization control problem, assuming a well-defined system transition function. This necessitates either a precise model or reliable forecasts. However, the challenges associated with these approaches arise from the difficulty of obtaining such precise models, especially in the presence of uncertainties.

In contrast, the model-free approaches offer an alternative that does not require an accurate model or prior knowledge of the environment.  Among these approaches, Deep Reinforcement Learning (DRL) stands out as a powerful technique for addressing sequential decision-making problems. In EV network, DRL allows for direct learning of optimal charging control behavior through interaction with the environment. Previous studies have utilized single-agent DRL techniques like Deep Deterministic Policy Gradient (DDPG) and Soft-Actor-Critic (SAC) for an individual EV or a group of EVs network, in which a single RL agent controls the charging power and charging scheduling of EVs during the charging phase. These studies operate under the assumption of full observability of the environment. This means that the DRL agent should have access to the local information of the EVs, such as their online consumption, arrival/departure time and their battery level. However, the EV owners may be unwilling to share these personal information with other owners. To tackle this concern, Multi-Agent RL (MARL) can be employed as an extension of single-agent RL. MARL is a powerful tool for distributed control, offering effective solutions to issues such as transformer overload and charging power control within EV networks.

In this paper, we model the charging control of an EVs network in residential sector. Our method utilizes a Partially Observable Markov Decision Process (POMDP) framework and incorporates a cutting-edge cooperative reinforcement learning technique called Multi-Agent Deep Deterministic Policy (MADDPG)~\cite{lowe2017multi}. Unlike previous works, we depart from the assumption of sharing global or local information among agents during execution. Instead, we adopt a centralized training-decentralized execution (CTDE) architecture that allows policies to leverage additional information for training purposes only, while still maintaining decentralized operation during execution. The only observation shared among agents in our framework is the electricity price. Once training is completed, only local actors are utilized during the execution phase, enabling effective and decentralized control that is applicable in cooperative settings. The main contributions of this article are as follows:

\begin{itemize}
    \item This paper presents a cooperative MARL framework for EVs network charging problem. The problem is formulated as POMDP, and the mathematical formulation for decentralized control has been proposed.
    \item We leverage the benefits of Centralized training decentralized execution framework to relax the assumption of observing the global network parameters and exchanging private information between agents.
    \item We examine the performance of our CTDE-oriented framework for a EV network. We provide the simulation results for up-to 10 agents and investigate the scalability of the proposed framework. 
\end{itemize}
The rest of this paper is organized as follows. Section~\ref{LitReview} provides the most recent related work. The network and system model formulation are described in Section~\ref{Background}. Section~\ref{Algorithm} introduced the proposed CTDE algorithm as well as the pricing scheme. Numerical results are presented in Section~\ref{results} and Section~\ref{conclusion} concludes the paper.

%% file: Conference_paper/Chapters/System_model2.tex
\section{Problem Formulation}\label{Background}

This paper presents a framework that automates energy management for an EV network on the demand-side, which can be utilized in DSM programs. The scenario we consider involves multiple end-users who share a common energy source, such as a generator or substation transformer connected to the grid. Each user is equipped with an energy consumption scheduler (ECS) installed in their smart meter. The smart meters interact automatically using a distributed algorithm to determine the optimal energy consumption for EVs. The optimization objective is to reduce the energy cost of the EV network while also minimizing the network Peak-to-Average(PAR). The primary aim of the network participants is to collaborate with each other to achieve the network's goals. It can also be seen that a pricing mechanism can provide the users with the incentives to cooperate. The price signal is the only information signal that broadcast to the end-users, and the users' aggregated demand is the only information signal send back to the utility company.
\begin{figure}[!t]
   \centering
    \includegraphics[scale=0.77, trim = 0.2cm 0.2cm 0.2cm 0.2cm, clip]{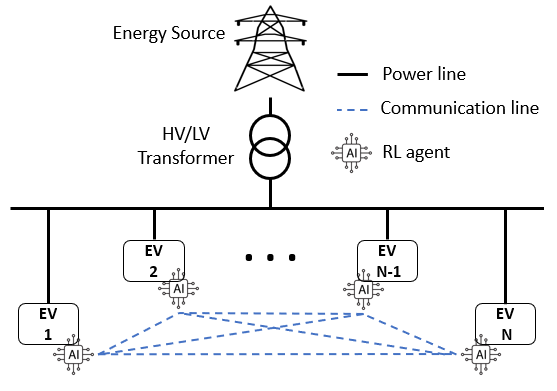}
    \caption{System model}
    \label{System Model}
    \vspace{-.1in}
\end{figure}

\subsection{System Model}
A block diagram of demand-side electric power distribution is shown in Fig.~\ref{System Model}. We assume that all the end-users are EV owners and are equipped with a smart meter and EV charging controller. The smart meters are all connected to the power line coming from the shared energy source, and they are also connected to each other through a communication channel. 

We model the distribution network in figure~\ref{System Model} as a graph $\mathcal{G} = (\mathscr{V},\mathscr{E})$, where $\mathscr{V} = \{0,1,...,N\}$ and $\mathscr{E} = \{1,2,...,M\}$ represent the set of nodes (users) and edges (branches), respectively. Node zero is considered as the connection to the shared energy source. For each user $i \in \mathscr{V}$, let $l_i^{h}$ denotes the total consumption of node $i$ at time $h$, where $ h \in \mathcal{H} = \{1,..., \mathscr{H}\}$. $\mathscr{H}$ denotes the last time of charging phase. Based on these definitions, $L_h = \sum\limits_{i \in \mathscr{V}}{l_i^h}$ represents the network total consumption at time $h$. For each user, let $\mathcal{X}_i$ denote the set of user appliances. Thus, the total consumption of the $i^{th}$ user is obtained as follows:
\begin{equation}
l_i^h = \sum\limits_{a \in \mathcal{X}_i}{x_{i,a}^h} = \sum\limits_{a \in \mathcal{X}_i \char`\\\{EV\}}{x_{i,a}^h} + x_{i,EV}^h
\end{equation}
where $x_{i,a}^h$ denotes the consumption of the appliance $a$ of the user $i$ at time $h$.
The focus of this paper is EV charging control. Thus, without lack of generality, we ignore the other appliances usages and the EV usage is the dominant term. Thus, hereinafter, by $l_i^h$ we mean $x_{i,EV}^h$. We are now ready to establish the optimization formulation for the EV network in our system model. Initially, we formulate the optimization of EV usage scheduling in a centralized manner, where a central controller has the ability to manage the scheduling problem. Subsequently, our focus shifts to reformulating the problem as a distributed optimization at the smart meter level, utilizing their Energy Control System (ECS) functionality.



\subsection{EV Network Centralized Minimization}
Considering the system model provided in Fig.~\ref{System Model}, the goal of an efficient EV charging controller can be twofold. Firstly, it aims to minimize the PAR to mitigate the risk of transformer overload during peak demand hours. Secondly, it aims to reduce the overall energy cost incurred by all end-users within the network. Here, we present the formulation of the EV charging scheduling problem aimed at minimizing the overall cost of the network, and we will illustrate that achieving goal one can be a consequence of accomplishing goal two. Hence, the EV charging scheduling problem is formulated as follows:
\begin{equation}\label{network_cost}
\min_{l_{i} \in \mathcal{X}_i} \sum\limits_{h=1}^{H} \mathbf{C}_h( \sum\limits_{i \in \mathscr{V}}{l_i^h}),
\end{equation}
where $\mathbf{C}_h$ denotes the network overall cost function at time step $h$.
Equation~\eqref{network_cost} can be solved in a centralized fashion using convex optimization programming techniques such as IPM~\cite{mohsenian2010autonomous}, but solving it this way requires a centralized controller that has access to the information of all users, which exposes the centralized node to potential privacy issues. To address this concern, it is more advantageous to define a distributed control approach, where the smart meters use their charging control functionality with minimum information exchange among the smart meters and energy source. In particular, the goal is to let each smart meter, with its EV charging optimizer, control the EV charging rate connected to it, according to the local information of the users such as arrival/departure time, EV battery level, and expected battery level. 

\subsection{EV Network Decentralized Optimization}
To define a distributed solution for the problem in~\eqref{network_cost}, let define $b_{i}^h$ as billing amount in dollars to be charged to the user $i$ by the utility at time slot $h$. At any given time, users are charged proportional to their total energy demand. This means:
\begin{equation}\label{ProportionalDemand}
    \frac{b_{i}^h}{b_{m}^h} = \frac{l_{i}^h}{l_{m}^h} \ \ \ \forall \ i,m \in \mathscr{V}.
\end{equation}

Using~\eqref{ProportionalDemand} for the total cost of the network from user $n$ perspective at any given time, we have:

\begin{equation}\label{extended_proportionalDemand}
    \sum\limits_{i \in \mathscr{V}}{b_{i}^h} = \sum\limits_{i \in \mathscr{V}}{\frac{{b_{m}^h} \times l_{i}^h}{l_{m}^h}} = \frac{b_{m}^h}{l_{m}^h} \sum\limits_{i \in {\mathscr{V}}}{l_{i}^h}.
\end{equation}

Together from \eqref{network_cost}, \eqref{ProportionalDemand}, and \eqref{extended_proportionalDemand} for each user we have:

\begin{align}\label{billing}
    b_{m}^h = \frac{l_{m}^h}{\sum\limits_{i \in \mathscr{V}}{l_{i}^h}}{\sum\limits_{i \in \mathscr{V}}{b_{i}^h}} = \frac{\kappa \times {l_{m}^h}}{\sum\limits_{i \in \mathscr{V}}{l_{i}^h}}{( \mathbf{C}_h( \sum\limits_{i \in \mathscr{V}}{l_i^h}))} =
    \nonumber \\
    \frac{\kappa \times {l_{m}^h}}{\sum\limits_{i \in \mathscr{V}}{l_{i}^h}}{( \mathbf{C}_h( {l_m^h}} + \sum\limits_{i \in \mathscr{V} \char`\\\{m\}}{l_{i}^h})),
\end{align}
where $\kappa$ is a constant coefficient. Equation \eqref{billing} illustrates that at any given time, the cost of the user $m$ not only depends on the local total consumption of the user $m$ ($l_m^h$), but also depends on other users' total consumption ($\sum\limits_{i \in \mathscr{V} \char`\\\{m\}}{l_{i}^h}$). 
Therefore, in distributed fashion, each agent aims to minimize its cost function as follows:
\begin{align}\label{NetwrokOptimization}
    \min_{l_{i}^h \in \mathcal{X}_i}
    \begin{aligned}[t]
      &\frac{\kappa \times {l_{i}^h}}{\sum\limits_{i \in \mathscr{V}}{l_{i}^h}}{( \mathbf{C}_h( {l_{i}^h}} + \sum\limits_{m \in \mathscr{V} \char`\\\{i\}}{l_{m}^h}))
   \end{aligned}
\end{align}

The optimization objective in~\ref{NetwrokOptimization} represents the total network cost from a single end-user's perspective. However, this objective is subject to various constraints that need to be taken into account. These constraints include \textbf{network-related factors} such as power balance, \textbf{physical EV limitations} like battery charging limits and rates, and battery maintenance considerations. Additionally, \textbf{EV owners' constraints}, such as arrival and departure times, expected battery level at departure, and remaining battery energy at arrival time, should also be considered.

Considering the corresponding constraints, the user $i$ can solve the problem in~\eqref{NetwrokOptimization} as long as it knows the total EV consumption of other users. It is important to note that user $i$ does not require the detail information of the EV consumption of other users within the network. Knowing the overall EV consumption of the network is enough. This problem has been solved in~\cite{mohsenian2010autonomous} using game-theory with two assumptions: (1) End-users are charged in proportion to their energy usage, which is unsuitable when dynamic pricing is implemented. (2) The daily energy consumption of all appliances, including EV, should be predetermined. The authors, in~\cite{yan2022cooperative}, solved this problem by relaxing these two assumptions, but they assume that  each EV is capable of observing the total network demand at any given time.
Nevertheless, in order to maintain privacy, it is not appropriate to share this information among users. Additionally, this is not information that can be accessed by users in the real world. Hence, how can the agent $i$ solve the problem defined in~\ref{NetwrokOptimization} locally, without having knowledge of the EV consumption of other users?

%% file: Conference_paper/Chapters/Conclusion.tex
\section{Conclusion}\label{conclusion}

In this study, we introduced an efficient decentralized framework for managing the charging phase of an Electric Vehicles (EVs) network. Our approach utilized a centralized training-decentralized execution deep deterministic policy gradient reinforcement learning framework. This framework allowed agents to gather additional information from other EVs exclusively during the training phase, while maintaining a fully decentralized strategy during the execution phase. We formulated the charging problem as a Partially Observable Markov Decision Process (POMDP). Furthermore, we conducted a comparative analysis between our proposed framework and a baseline approach where independent DDPG agents individually solve their local charging problems without any information from other agents, even during the training phase. Our simulation results demonstrate that with coordination among agents, the overall network cost and average electricity price decrease, leading to reduced individual costs. Additionally, in terms of the scalability our results indicate that the CTDE outperform the independent DDPG when the number of agents increases.